\documentclass{article}
\usepackage{amsmath,amssymb,cite,epsfig,graphicx,fancyhdr}
\begin{document}
\lhead {Solution hierarchies for the Painlev\'e IV equation}
\rhead{D. Berm\'udez, D.J. Fern\'andez C.}

\title {Solution hierarchies for the Painlev\'e IV equation}
\author{David Berm\'udez\footnote{{\it email:} dbermudez@fis.cinvestav.mx},\, David J. Fern\'andez C.\footnote{{\it email:} david@fis.cinvestav.mx} \\ \vspace{2mm}
{\sl Departamento de F\'{\i}sica, Cinvestav, A.P. 14-740, 07000 M\'exico D.F., Mexico}}

\date{}
\maketitle

\begin{abstract}
We will obtain real and complex solutions to the Painlev\'e IV equation through supersymmetric quantum mechanics. The real solutions will be classified into several hierarchies, and a similar procedure will be followed for the complex solutions.
\end{abstract}

\section{Introduction}

The Painlev\'e equations can be seen as the nonlinear analogues of the classical linear equations associated to the well known special functions \cite{VS93,Adl94}. They have been identified as the most important non-linear ordinary differential equations \cite{TKS91}. Although discovered from strictly mathematical considerations, nowadays they are widely used to describe several physical phenomena \cite{AC92}. In particular, the Painlev\'e IV equation ($P_{IV}$) is relevant in fluid mechanics, non-linear optics, and quantum gravity \cite{Win92}.

On the other hand, since its birth supersymmetric quantum mechanics (SUSY QM) catalyzed the study of exactly solvable Hamiltonians and gave a new insight into the algebraic structure characterizing these systems. Historically, the essence of SUSY QM was developed first as Darboux transformation in mathematical physics \cite{MS91} and as factorization method in quantum mechanics \cite{IH51,mi84}. Moreover, through SUSY QM one can obtain quantum systems described by second-order polynomial Heisenberg algebras (PHA), whose Hamiltonians have the standard Schr\"odinger form and their differential ladder operators are of third order. It has been shown that there is a connection between these systems and solutions $g(x;a,b)$ of $P_{IV}$ \cite{Adl94}.

The $P_{IV}$ solutions can be grouped into several hierarchies, according to the family of special functions they are related with. This classification can be easily done for the class of real solutions \cite{BF11}, but it can be as well performed for the recently found complex solutions \cite{BF11a}, which is our aim here. To do that, we have arranged this paper as follows: in Section 2 we shall present the general framework of SUSY QM and PHA. In the next Section we will generate the real and complex solutions to $P_{IV}$; then, in Section 4 we will study the real solution hierarchies and we shall analyze the domain of the parameter space $(a,b)$ where they are to be found. In Section 5 we do the same for the complex solution. We present our conclusions in Section 6.

\section{General framework of SUSY QM and PHA}

In the $k$-th order SUSY QM one starts from a given solvable Hamiltonian
\begin{eqnarray}
&& H_0  = -\frac12 \frac{d^2}{d x^2} + V_0(x),
\end{eqnarray}
and generates a chain of first-order intertwining relations \cite{AIS93,MRO04,Fer10}
\begin{eqnarray}
&& H_j A_j^{+} = A_j^{+} H_{j-1}, \quad  H_{j-1}A_j^{-} = A_j^{-}H_j,\label{susy1} \\
&& \hskip-1.2cm H_j = -\frac12 \frac{d^2}{d x^2} + V_j(x),\label{susy2}\quad
A_j^{\pm} = \frac{1}{\sqrt{2}}\left[\mp \frac{d}{d x} + \alpha_j(x,\epsilon_j)\right], \quad j = 1,\dots,k.\label{susy3}
\end{eqnarray}
By plugging Eqs.~\eqref{susy2} into Eq.~\eqref{susy1} we obtain
\begin{eqnarray}
&& \hskip-1cm \alpha_j'(x,\epsilon_j) + \alpha_j^2(x,\epsilon_j) = 2[V_{j-1}(x) - \epsilon_j],  \label{rei} \quad
V_{j}(x) = V_{j-1}(x) - \alpha_j'(x,\epsilon_j). \label{npi}
\end{eqnarray}
We are interested in the final Riccati solution $\alpha_{k}(x,\epsilon_{k})$, which turns out to be determined either by $k$ solutions $\alpha_1(x,\epsilon_j)$ of the initial Riccati equation
\begin{equation}
\alpha_1'(x,\epsilon_j) + \alpha_1^2(x,\epsilon_j) = 2 [V_0(x) - \epsilon_j], \quad j=1,\dots,k,
\end{equation}
or by $k$ solutions $u_j \propto \exp(\int \alpha_1(x,\epsilon_j)dx)$ of the associated Schr\"odinger equation
\begin{eqnarray}
&& H_0 u_j = - \frac12 u_j'' + V_0(x)u_j = \epsilon_j u_j, \quad j=1,\dots,k. \label{usch}
\end{eqnarray}

Thus, there is a pair of $k$-th order operators interwining the initial $H_0$ and final Hamiltonians $H_k$, namely,
\begin{eqnarray}
H_k B_k^{+} = B_k^{+} H_0, \ H_0 B_k^{-} = B_k^{-} H_k, \
B_k^{+} = A_k^{+}\dots A_1^{+}, \ B_k^{-} = A_1^{-}\dots A_k^{-}.
\end{eqnarray}
The normalized eigenfunctions $\psi_n^{(k)}$ of $H_k$, associated to the eigenvalues $E_n$, and the $k$ additional eigenstates $\psi_{\epsilon_j}^{(k)}$ associated to the eigenvalues $\epsilon_j$ which are annihilated by $B_k^{-}$ ($j=1,\dots ,k$), are given by \cite{BF11,FH99}:
\begin{eqnarray}
&& \hskip-1.2cm \psi_n^{(k)} \! = \! \frac{B_k^{+}\psi_n}{\sqrt{(E_n-\epsilon_1)\dots (E_n-\epsilon_k)}}, \label{psin} \
\psi_{\epsilon_j}^{(k)} \! \propto \! \frac{W(u_1,\dots , u_{j-1},u_{j+1},\dots , u_k)}{W(u_1,\dots , u_k)}. \label{psie}
\end{eqnarray}
Note that, in this formalism the obvious restriction $\epsilon_j < E_0=1/2$ naturally arises since we want to avoid singularities in $V_k(x)$.

On the other hand, a $m$-th order PHA is a deformation of the Heisenberg-Weyl algebra of kind \cite{FH99,fnn04,CFNN04}:
\begin{eqnarray}
 && \hskip-1.3cm [H,L^\pm]  = \pm L^\pm , \quad
[L^-,L^+] 	\equiv Q_{m+1}(H+1) - Q_{m+1}(H) = P_m(H) , \\
&& Q_{m+1}(H) = L^+ L^- = \left(H - \mathcal{E}_1\right) \dots \left(H - \mathcal{E}_{m+1}\right) ,
\end{eqnarray}
where $P_m(x)$ is a polynomial of order $m$ in $x$ and $\mathcal{E}_i$ are the zeros of $Q_{m+1}(H)$, which correspond to the energies associated to the extremal states of $H$.

Now, in the differential representation of the second-order PHA ($m=2$), $L^+$ is a third-order differential ladder operator, chosen by simplicity as \cite{acin00}:
\begin{eqnarray}
& \hskip-0.7cm L^+   = L_1^+ L_2^+ , \
L_1^+ = \frac{1}{\sqrt{2}}\left[-\frac{d}{d x} + f(x) \right], \
L_2^+ = \frac12[\frac{d^2}{d x^2} + g(x)\frac{d}{d x} + h(x)].
\end{eqnarray}
These operators satisfy the following relationships:
\begin{eqnarray}
& H L_1^+ = L_1^+ (H_{\rm a} + 1), \quad H_{\rm a} L_2^+ = L_2^+ H \quad
 \Rightarrow \quad  [H,L^+] = L^+,
\end{eqnarray}
$H_{\rm a}$ being an auxiliary Schr\"odinger Hamiltonian. Using the standard first and second-order SUSY QM gives rise to
\begin{align}
& f = x + g, \label{fdependg} \qquad
h = - x^2 + \frac{g'}{2} - \frac{g^2}{2} - 2xg + a, \\
& V = \frac{x^2}2 - \frac{g'}2 + \frac{g^2}2 + x g + \mathcal{E}_1 -
\frac12 , \label{Vpivs} \\
& g'' = \frac{g'^2}{2g} + \frac{3}{2} g^3 + 4xg^2 + 2\left(x^2 - a \right) g + \frac{b}{g}.  \label{painleveiv}
\end{align}
The last is the Painlev\'e IV equation ($P_{IV}$) with parameters
\begin{equation}
a =\mathcal{E}_2 + \mathcal{E}_3-2\mathcal{E}_1 -1,\quad b = - 2(\mathcal{E}_2 - \mathcal{E}_3)^2.\label{abe}
\end{equation}
If the $\mathcal{E}_i, \ i=1,2,3$ are real, we will obtain real parameters $a,b$ for Eq.~\eqref{painleveiv}.

\section{Real and complex solutions to $P_{IV}$ with real parameters}

It is well known that the first-order SUSY partner Hamiltonians of the harmonic oscillator are naturally described by second-order PHA, which are connected with $P_{IV}$. Furthermore, there is a theorem stating the conditions for the hermitian higher-order SUSY partners Hamiltonians of the harmonic oscillator to have this kind of algebras (see \cite{BF11}). The main requirement is that the $k$ Schr\"odinger seed solutions have to be connected in the way
\begin{eqnarray}
u_j=(a^{-})^{j-1}u_1, \qquad \label{us}
\epsilon_j=\epsilon_1-(j-1),  \qquad j=1,\dots , k,
\end{eqnarray}
where $a^{-}$ is the standard annihilation operator of $H_0$ so that $u_1$ is the only free seed.

If $u_1$ is a real solution of Eq.~\eqref{usch} without zeros, associated to a real factorization energy $\epsilon_1$ such that $\epsilon_1<E_0=1/2$, then all $u_j$ are also real and, consequently, the solutions to $P_{IV}$ are also real. On the other hand, if we use the formalism as in \cite{BF11} with $\epsilon_1 > E_0$, we would obtain only singular SUSY transformations. In order to avoid this we will instead employ complex SUSY transformations. The simplest way to implement them is to use a complex linear combination of the two standard linearly independent real solutions which, up to an unessential factor, leads to the following complex solutions depending on a complex constant $\Lambda=\lambda + i \kappa$ ($\lambda, \kappa \in \mathbb{R}$) \cite{ACDI99}:
\begin{eqnarray}
& u(x;\epsilon ) =  e^{-x^2/2}\left[ {}_1F_1\left(\frac{1-2\epsilon}{4},\frac12;x^2\right) + x \, \Lambda \, {}_1F_1\left(\frac{3-2\epsilon}{4},\frac32;x^2\right)\right], \label{u1}
\end{eqnarray}
where $_1F_1$ is the {\it confluent hypergeometric function}. The results for the real case \cite{JR98} are obtained by making $\kappa=0$ and expressing $\Lambda=\lambda$, with $\nu \in \mathbb{R}$, as
\begin{eqnarray}
& \Lambda = \lambda= 2 \nu\frac{\Gamma(\frac{3 - 2\epsilon}{4})}{\Gamma(\frac{1-2\epsilon}{4})}. \label{nu}
\end{eqnarray}

Note that the extremal states of $H_{k}$ and their corresponding energies are given by
\begin{align}
\psi_{\mathcal{E}_1} \propto \frac{W(u_1,\dots,u_{k-1})}{W(u_1,\dots,u_k)}, & \quad \mathcal{E}_1 = \epsilon_k = \epsilon_1 - (k - 1), \label{edo1}\\
\psi_{\mathcal{E}_2} \propto B_k^+ e^{-x^2/2}, & \quad \mathcal{E}_2 = \frac{1}{2}, \label{edo2}\\
\psi_{\mathcal{E}_3} \propto B_k^+ a^{+} u_1, & \quad \mathcal{E}_3 = \epsilon_1 + 1. \label{edo3}
\end{align}
Recall that all the $u_j$ satisfy Eq.~\eqref{us} and $u_1$ corresponds to the general solution given in Eq.~\eqref{u1}.

Hence, through this formalism we will obtain a $k$-th order SUSY partner potential $V_k(x)$ of the harmonic oscillator and a $P_{IV}$ solution $g_k(x;\epsilon_1)$, both of which can be chosen real or complex, in the way
\begin{align}
V_k(x) &= \frac{x^2}2 - \{\ln [W(u_1,\dots,u_k)]\}'' , \\
g_k(x;\epsilon_1) &= - x - \{\ln[\psi_{\mathcal{E}_1}(x)]\}'. \label{solg}
\end{align}

For $k=1$, the first-order SUSY transformation and Eq.~\eqref{solg} lead to what is known as \emph{one-parameter solutions} to $P_{IV}$, due to the restrictions imposed by Eq.~\eqref{abe} onto the parameters $a,b$ of $P_{IV}$ which make them both depend on $\epsilon_1$ \cite{BCH95}. For this reason, this family of solutions cannot be found in any point of the parameter space $(a,b)$, but only in the subspace defined by the curve $\{\left( a(\epsilon_1), b(\epsilon_1)\right),\ \epsilon_1 \in \mathbb{R}\}$ consistent with Eqs.~\eqref{abe}. Then, by increasing the order of the transformation to an arbitrary integer $k$, we will expand this subspace for obtaining $k$ different families of one-parameter solutions. This procedure is analogous to iterated auto-B\"acklund transformations \cite{RS82}. Also note that by making cyclic permutations of the indices of the three energies $\mathcal{E}_i$ and the corresponding extremal states of Eqs.~(\ref{edo1}-\ref{edo3}) (when they have no nodes), we expand the solution families to three different sets, defined by
\begin{align}
a_{1}=-\epsilon_1 + 2k -\frac{3}{2}, \quad & b_{1}=-2\left(\epsilon_1+\frac{1}{2}\right)^{2}, \label{ab1}\\
a_2= 2\epsilon_1 -k, \quad & b_2=-2k^2, \\
a_3=-\epsilon_1-k-\frac{3}{2}, \quad & b_3=-2\left(\epsilon_1 - k +\frac{1}{2}\right)^2,
\end{align}
where we have added an index corresponding to the extremal state given by Eqs.~(\ref{edo1}-\ref{edo3}). Therefore we obtain three different solution families of $P_{IV}$ through Eqs.~(\ref{u1}-\ref{solg}). The first family includes non-singular real and complex solutions, while the second and third ones can give just non-singular strictly complex solutions, due to singularities appearing in the real case.

\section{Real solution hierarchies}
The solutions $g_k(x;\epsilon_1)$ of the Painlev\'e IV equation can be classified according to the explicit functions they depend on \cite{BCH95}. In the real case, see Eqs.~(\ref{u1},\ref{solg}) with the condition given in Eq.~\eqref{nu}, the solutions are expressed in terms of the confluent hypergeometric function $_1F_1$, although for specific values of the parameter $\epsilon_1$ they can be reduced to the {\it error function} $\text{erf}(x)$. Moreover, for particular parameters $\epsilon_1$ and $\nu_1$, they simplify further to rational solutions.

Let us remark that we are interested in non-singular SUSY partner potentials and the corresponding non-singular solutions of $P_{IV}$. Note that the same set of real solutions to $P_{IV}$ can be obtained through inverse scattering techniques \cite{AC92} (compare the solutions of \cite{BCH95} with those of \cite{BF11}).

\subsection{Confluent hypergeometric function hierarchy}
In general, the solutions of $P_{IV}$ are expressed in terms of two confluent hypergeometric functions. For example, let us write down the explicit formula for $g_1(x;\epsilon_1)$ in terms of the parameters $\epsilon_1,\, \nu_1$ (with $\epsilon_1 < 1/2$ and $\vert \nu_1 \vert < 1$ to avoid singularities):

\begin{align}\hspace{-8.3pt}\displaystyle
g_1(x,\epsilon_1)=&  \frac{2\nu_1\Gamma\left(\frac{3-2\epsilon_1}{4}\right) \left[3\,{}_1F_1\left(\frac{3-2\epsilon_1}{4},\frac{3}{2};x^2\right)-(2\epsilon_1 +3)x^2\,{}_1F_1\left(\frac{3-2\epsilon_1}{4},\frac{5}{2};x^2\right) \right]}
{3 \Gamma\left(\frac{1-2\epsilon_1}{4}\right)\,{}_1F_1\left(\frac{1-2\epsilon_1}{4},\frac{1}{2};x^2\right) + 6\nu_1 x \Gamma\left(\frac{3-2\epsilon_1}{4}\right)\,{}_1F_1(\frac{3-2\epsilon_1}{4},\frac{3}{2};x^2) }\nonumber\\
&\hskip-1.3cm +\frac{3x(2\epsilon_1 +1)\Gamma\left(\frac{1-2\epsilon_1}{4}\right)\,{}_1F_1\left(\frac{1-2\epsilon_1}{4},\frac{3}{2};x^2\right)}
{3 \Gamma\left(\frac{1-2\epsilon_1}{4}\right)\,{}_1F_1\left(\frac{1-2\epsilon_1}{4},\frac{1}{2};x^2\right) + 6\nu_1 x \Gamma\left(\frac{3-2\epsilon_1}{4}\right)\,{}_1F_1(\frac{3-2\epsilon_1}{4},\frac{3}{2};x^2) }. \label{g1}
\end{align}
The explicit analytic formulas for higher-order solutions $g_k(x;\epsilon_1)$ can be obtained from expression \eqref{solg}, and they have a similar form as in Eq.~\eqref{g1}.

\subsection{Error function hierarchy}
It is interesting to analyze the possibility of reducing the explicit form of the $P_{IV}$ solution to the error function. To do that, let us fix the factorization energy in such a way that any of the two hypergeometric series of Eq.~\eqref{u1} reduces to that function. This can be achieved for $\epsilon_1 = -(2m+1)/2$, with $m\in\mathbb{N}$. By defining $\varphi_{\nu_1}(x)\equiv \sqrt{\pi}e^{x^2}[1+\nu_1\, \text{erf}(x)]$, we can write down simple expressions for $g_k(x,\epsilon_1)$ for some specific parameters $k$ and $\epsilon_1$:
\begin{align}
g_1(x;-5/2)&=\frac{4[\nu_1 + x\varphi_{\nu_1}(x)]}{2\nu_1 x +(1+2x^2)\varphi_{\nu_1}(x)},\label{erf2}\\
g_2(x;-1/2)&=\frac{4\nu_1[\nu_1 + 6x\varphi_{\nu_1}(x)]}{\varphi_{\nu_1}(x)[\varphi_{\nu_1}^2(x) -2\nu_1 x \varphi_{\nu_1}(x) -2\nu_1^2]}.
\end{align}

\subsection{Rational hierarchy}
\begin{figure}
\begin{center}
\includegraphics[scale=0.4]{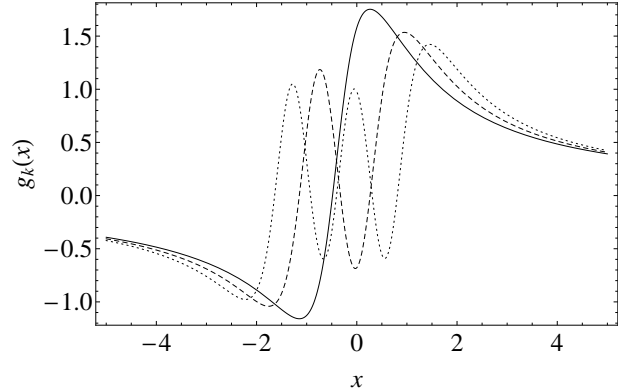}
\end{center}
\vspace{-5mm}
\caption{The $P_{IV}$ solutions given by Eqs.~(\ref{rg1}--\ref{rg3}).} \label{realsol}
\end{figure}

Now, let us look for the restrictions needed to reduce the explicit form of Eq.~\eqref{solg} to non-singular rational solutions. To achieve this, once again the factorization energy $\epsilon_1$ has to be a negative half-integer, but depending on the $\epsilon_1$ taken, just one of the two hypergeometric functions is reduced to a polynomial. Thus, we need to choose additionally the parameter $\nu_1=0$ or $\nu_1\rightarrow\infty$ to keep the appropriate hypergeometric function. However, $u_1$ have a zero at $x=0$ when $\nu_1\rightarrow \infty$, which will produce one singularity for the corresponding $P_{IV}$ solution. Hence, we should make $\nu_1 = 0$ and $\epsilon_1 = -(4m+1)/2$ with $m\in\mathbb{N}$. Departing from Schr\"odinger solutions with these $\nu_1$, $\epsilon_1$ we get some explicit expressions for the $g_k(x;\epsilon_1)$ of the rational hierarchy:
\begin{align}
g_1(x;-5/2)&=\frac{4 x}{1 + 2 x^2},\label{rg1}\\
g_2(x;-5/2)&= -\frac{4 x}{1 + 2 x^2} + \frac{16 x^3}{3 + 4 x^4},\\
g_3(x;-5/2)&= -\frac{16 x^3}{3 + 4 x^4}+\frac{12(3x+4x^3+4x^5)}{9+18x^2-12x^4+8x^6},\label{rg3}
\end{align}
which are plotted in Fig.~\ref{realsol}.

\subsection{First kind modified Bessel function hierarchy}
Another interesting case associated to a special function arises for $\epsilon_1=-m$, $m\in\mathbb{N}$, which leads to the modified Bessel function of first kind. We write down an example of one solution belonging to such a hierarchy:
\begin{equation}
g_1(x;0)=\frac{\nu_1(1-x^2)I_{\frac{1}{4}}\left(\frac{x^2}{2}\right)+x^2\left[-I_{-\frac{1}{4}}\left(\frac{x^2}{2}\right)+I_{\frac{3}{4}}\left(\frac{x^2}{2}\right)+\nu_1 I_{\frac{5}{4}}\left(\frac{x^2}{2}\right)\right]}{x\left[I_{-\frac{1}{4}}\left(\frac{x^2}{2}\right)+\nu_1I_{\frac{1}{4}}\left(\frac{x^2}{2}\right)\right]}.
\end{equation}

\begin{figure}
\begin{center}
\includegraphics[scale=0.4]{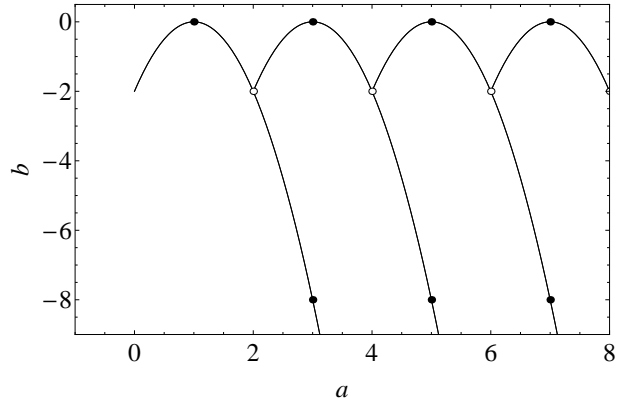}
\end{center}
\vspace{-5mm}
\caption{Parameter space for real $P_{IV}$ solutions. The lines represent solutions of the confluent hypergeometric function hierarchy, the black dots of the error function hierarchy, and the white dots of the rational and error function hierarchies.} \label{realpara}
\end{figure}

\section{Complex solution hierarchies}
Let us study the complex solutions subspace, i.e., we use the complex linear combination of Eq.~\eqref{u1} and the associated $P_{IV}$ solution of Eq.~\eqref{solg}. This allows the use of seeds $u_1$ with $\epsilon_1 \geq 1/2$ but without producing singularities. Moreover, the complex case is richer than the real one, since all three extremal states of Eqs.~\eqref{edo1}-\eqref{edo3} lead to non-singular complex $P_{IV}$ solution families.

\subsection{Confluent hypergeometric hierarchy}
As in the real case, in general the solutions of $P_{IV}$ are expressed in terms of two confluent hypergeometric functions. In particular,
the explicit formula for the first family $g_1(x;\epsilon_1)$ in terms of the parameters $\epsilon_1,\, \Lambda$ is given by
\begin{align}\displaystyle
g_1(x,\epsilon_1)=&  \frac{\Lambda\left[3\,{}_1F_1\left(\frac{3-2\epsilon_1}{4},\frac{3}{2};x^2\right)-(2\epsilon_1+3)x^2 {}_1F_1\left(\frac{3-2\epsilon_1}{4},\frac{5}{2};x^2\right)\right]}
{3\,{}_1F_1\left(\frac{1-2\epsilon_1}{4},\frac{1}{2},x^2\right)+\Lambda \,x\,{}_1F_1\left(\frac{3-2\epsilon_1}{4},\frac{3}{2},x^2\right)}\nonumber\\
&+\frac{-3x(2\epsilon_1 + 1)\,{}_1F_1\left(\frac{1-2\epsilon_1}{4},\frac{3}{2};x^2\right)}
{3\,{}_1F_1\left(\frac{1-2\epsilon_1}{4},\frac{1}{2},x^2\right)+\Lambda\,x\,{}_1F_1\left(\frac{3-2\epsilon_1}{4},\frac{3}{2},x^2\right)}. \label{g1complex}
\end{align}
Once again, for all families the explicit analytic formulas for the higher-order solutions $g_k(x;\epsilon_1)$ can be obtained through the formula \eqref{solg}.

\subsection{Error function hierarchy}
If we choose the parameter $\epsilon_1=-(2m+1)/2$ with $m\in\mathbb{N}$, as in the real case, we obtain the error function hierarchy. In terms of the auxiliary function $\phi_{\Lambda}=\text{e}^{x^2}[4 + \Lambda \pi^{1/2}\text{erf}(x)]$, a solution from the third family is written as:
\begin{equation}
g_1(x;-5/2)=\frac{4\Lambda + 4x\phi_{\Lambda}(x)}{2\Lambda x+(1+2x^2)\phi_{\Lambda}(x)}.
\end{equation}
\subsection{Imaginary error function hierarchy}
\begin{figure}
\begin{center}
\includegraphics[scale=0.4]{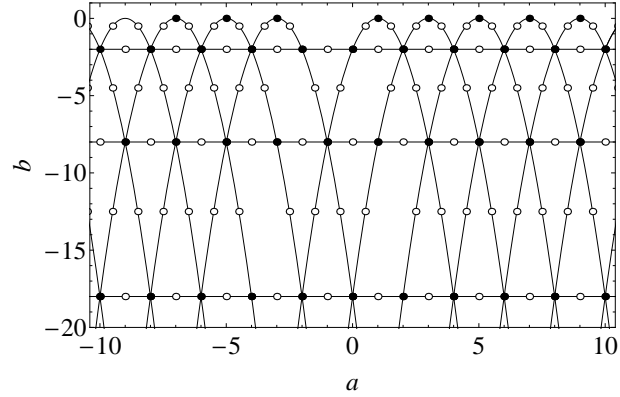}
\end{center}
\vspace{-5mm}
\caption{Parameter space for complex solution hierarchies. The lines correspond to the confluent hypergeometric function, the black dots to the error function or the imaginary error function, and the white dots to the first kind modified Bessel function.} \label{complexpara}
\end{figure}

Different to the real case, now we can use $\epsilon_1\geq 1/2$, giving place to more solution families. This is clear by comparing the real and  complex parameter spaces of solutions from Fig.~\ref{realpara} and Fig.~\ref{complexpara}. By defining a new auxiliary function $\phi i_{\Lambda}=\text{e}^{-x^2}[4 + \Lambda \pi^{1/2}\text{erfi}(x)]$, where $\text{erfi}(x)$ is the {\it imaginary error function}, we can write down an explicit solution from the third family
\begin{equation}
g_1(x;5/2)=\frac{4\Lambda(1-x^2)+2x(-3+2x^2)\phi i_{\Lambda}(x)}
{2\Lambda x+(1-2x^2)\phi i_{\Lambda}(x)}.
\end{equation}

\subsection{First kind modified Bessel function hierarchy}
\begin{figure}
\begin{center}
\includegraphics[scale=0.4]{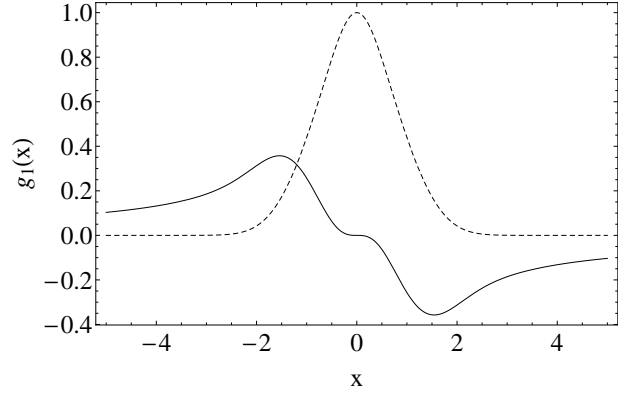}
\end{center}
\vspace{-5mm}
\caption{Real (solid curve) and imaginary (dashed curve) parts of a complex solution to $P_{IV}$. The plot corresponds to $k=1$, $\epsilon_1=0$, $\lambda=0$, and $\kappa=1$.} \label{complexsol}
\end{figure}

Let us write down an example of the solution of this hierarchy for $\lambda=0$, $\kappa=1$, $\Lambda = i$, i.e., $u_1$ is a purely imaginary linear combination of the two standard real solutions associated to $\epsilon_1 = 0$:
\begin{equation}
g_1(x;0)=\frac{x\Gamma\left(\frac{3}{4}\right)\left[I_{\frac{3}{4}}\left(\frac{x^2}{2}\right)-I_{-\frac{1}{4}}\left(\frac{x^2}{2}\right)\right]+2 i x \Gamma\left(\frac{5}{4}\right)\left[I_{-\frac{3}{4}}\left(\frac{x^2}{2}\right)-I_{\frac{1}{4}}\left(\frac{x^2}{2}\right)\right]}
{\Gamma\left(\frac{3}{4}\right)I_{-\frac{1}{4}}\left(\frac{x^2}{2}\right)+2 i \Gamma\left(\frac{5}{4}\right)I_{\frac{1}{4}}\left(\frac{x^2}{2}\right)}.
\end{equation}
Its real and imaginary parts are plotted in Fig.~\ref{complexsol}.

\section{Conclusions}
In this paper we have discussed a general method to obtain real and complex solutions of Painlev\'e IV equation by using SUSY QM, which is closely related to the factorization method. Through this scheme we have shown that real factorization energies can be used to obtain $P_{IV}$ solutions with real parameters $a,b$. We have shown the existence of more solutions in the complex case than in the real one by studying in detail the parameter space $(a,b)$.

We have classified the solutions into hierarchies arising both in the real and in the complex cases. Both classifications became very similar, except for a hierarchy which cannot be obtained in the real case. A further study of the Painlev\'e IV equation with complex parameters is currently under study.

\section*{Acknowledgement}
The authors acknowledge the financial support of Conacyt, project 152574. DB also acknowledges the Conacyt  PhD scholarship 219665.

\end{document}